\documentclass[aps,prl,preprint,amsmath,amssymb,groupedaddress]{revtex4-1}
\usepackage{graphicx}
\usepackage{dcolumn}
\usepackage{bm}

\begin{document}


\title{Chiral Surface Waves for Enhanced Circular Dichroism}


\author{Giovanni Pellegrini}
\email[]{giovanni.pellegrini@polimi.it}
\author{Marco Finazzi}
\author{Michele Celebrano}
\author{Lamberto Du\`{o}}
\author{Paolo Biagioni}
\email[]{paolo.biagioni@polimi.it}

\affiliation{Dipartimento di Fisica, Politecnico di Milano, Piazza Leonardo da Vinci 32, 20133 Milano, Italy}


\date{\today}

\begin{abstract}
We present a novel chiral sensing platform that combines a one-dimensional photonic
crystal design with a birefringent surface defect. The platform sustains simultaneous transverse
electric and transverse magnetic surface modes, which are exploited to generate chiral surface
waves. The present design provides homogeneous and superchiral fields of both handednesses over
arbitrarily large areas in a wide spectral range, resulting in the enhancement of the circular
dichroism signal by more than two orders of magnitude, thus paving the road toward the successful
combination of surface-enhanced spectroscopies and electromagnetic superchirality.
\end{abstract}


\maketitle

Chiral objects are three dimensional bodies characterized by being geometrically distinct from their
mirror image. Objects of opposite chirality, called enantiomers, share most of their physical
properties and their chiral nature emerges only when they interact with a chiral environment. This
geometrical feature, which plays a crucial role in many biochemical processes, is ubiquitous in
biomolecules, the most outstanding example being DNA~\cite{Meierhenrich_2008,Fasman_1996}. As a
consequence, the analysis and separation of chiral molecules has lately gained increasing importance
in the pharmaceutical and biochemical industry. Circular dichroism (CD) spectroscopy, i.e. the
measurement of the differential absorption between left (L) and right (R) circularly polarized light
(CPL), is one of the most relevant tools for enantiomer discrimination. However, CD signals are
usually extremely weak if compared to those of standard achiral absorption spectroscopies, which
makes the analysis of small amounts of chiral analytes extremely
challenging~\cite{barron_molecular_2004,Hendry_2010,Tang_2010,Tang_2011}.

In this context, novel approaches have been proposed over the last few years to enhance the CD
signal by tailoring the chiral properties of the probing electromagnetic
field~\cite{Tang_2010,Tang_2011,Sch_ferling_2012,Sch_ferling_2014,Nesterov_2016} through the control
of the associated optical chirality $C$~\cite{Tang_2010}, which is defined as:
\begin{equation}
C=-\frac{\varepsilon_{\mathrm{0}}\omega}{2}\mathrm{Im}(\mathbf{E^{*}}\cdot\mathbf{B}).
\label{eq:chirality}
\end{equation}
This quantity is a time-even pseudoscalar, i.e. it behaves as a chiral object under mirror
inversion, and determines the degree of asymmetry in the absorption rate of a chiral molecule
between L and R circularly polarized light in the dipolar approximation. Circularly polarized
plane waves are a common example of a chiral electromagnetic field, with their optical chirality for
a unitary intensity wave expressed as $C_{\mathrm{CPL}}^{\mathrm{L,R}} = \pm
\varepsilon_{\mathrm{0}}\omega/2$~\cite{Tang_2010}. Likewise, a field distribution where $C$ is
beyond the $C_{\mathrm{CPL}}^{\mathrm{L,R}}$ value is dubbed `superchiral'.

In this framework, an ideal chiral sensing platform should be able to
(i) provide uniform superchiral optical fields ($|C|>|C_{\mathrm{CPL}}^{\mathrm{L,R}}|$) over large
areas, (ii) generate optical chiralities of both handednesses upon reversing the polarization state
of the incident field, and (iii) work at wavelengths ranging from the near-UV to the IR depending on
the platform design, with a particular attention to the high energy range of the spectrum, below
400~nm, where most electronic molecular transitions
occur~\cite{Abdulrahman_2012,Sch_ferling_2016,Nesterov_2016}.

Plasmonic nanostructures have been widely proposed and employed as a viable solution for the
enhancement of the optical chiral response of
biomolecules~\cite{Hendry_2010,Abdulrahman_2012,Govorov_2010,Nesterov_2016,Govorov_2012,Hendry_2012,
Sch_ferling_2012,Sch_ferling_2012a,Valev_2013,Lu_2013,
Frank_2013,Liu_2014,Valev_2014,Sch_ferling_2014,Sch_ferling_2016,Tullius_2015,Finazzi_2015}.
However, despite the recent results, to date chiral plasmonic sensing does not meet all the
above criteria simultaneously. In particular, superchiral optical fields are spatially confined to
the so-called plasmonic `hot spots' and large homogeneous optical chiralities are usually obtained
only for chiral plasmonic nanostructures, thus hindering the possibility of handedness switching
upon reversal of the incident polarization state. Indeed, the utmost importance of molecular
spectroscopy in the blue and near-UV energy range, which is not accessible with standard Au
plasmonics, is pushing for the adoption of novel plasmonic materials~\cite{Knight_2014}.

Uniform surface-enhanced fields for sensing, which are not found in substrates based on
nanoparticles, are instead obtained when surface plasmon polaritons or Bloch surface waves are
employed~\cite{Yeh_1977,maier_2007,Descrovi_2015}. However, surface plasmon polaritons, which are
propagating longitudinal oscillations of the electron plasma at the surface of a metal, are
intrinsically transverse magnetic (TM) surface modes, therefore generating local fields that carry
no optical chirality, in analogy with linearly polarized plane waves. Bloch surface waves, on the
other side, are surface waves created by a semi-infinite one dimensional photonic crystal (1DPC),
usually terminated with a surface defect, and support both transverse electric (TE) and TM modes.
However, the two dispersion relations of such modes do not overlap in ordinary conditions
and therefore the TE and TM waves cannot be excited simultaneously, again preventing the creation of
chiral surface waves (CSWs).

To address all these issues, we introduce a novel chiral sensing platform based on the combination
of a 1DPC with a properly engineered anisotropic surface defect. Electrodynamics calculations show
that such a platform, which is well within the modern fabrication capabilities, supports CSWs
originating from the coherent superposition of TE and TM surface modes, providing (i) homogeneous
and (ii) switchable superchiral fields over arbitrarily large areas and wide spectral ranges.
Additionally, (iii) the use of standard dielectric materials allows moving the operation wavelength
toward the high-energy blue end of the spectrum.
\begin{figure}[t!]
\includegraphics{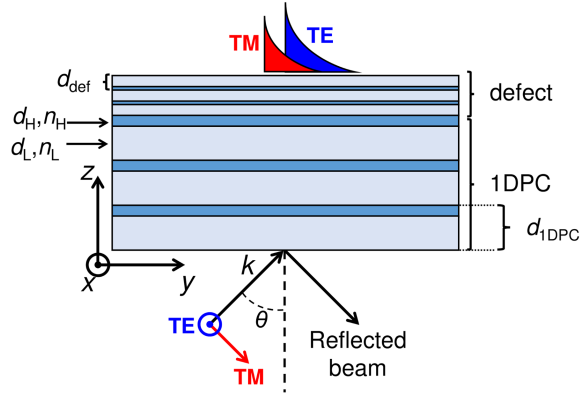}
\caption{\label{fig:1} (Color online) A schematic representation of the 1DPC sensing platform, with
the incoming TE and TM linearly polarized waves represented at the bottom, and the corresponding TE
and TM surface modes at the top. Crystal and defect periods are defined as $d_{\mathrm{1DPC}}$ and
$d_{\mathrm{def}}$ respectively, while $d_{\mathrm{H,L}}$ stands for the thickness of the high and
low refractive index materials in the 1DPC.}
\end{figure}
The design of the chiral sensing platform is based on two fundamental considerations regarding the
optical properties of a semi-infinite 1DPC terminated with a surface defect. Specifically, (i) 1DPCs
support both TE and TM surface modes, and (ii) the slopes of the respective dispersion relations are
a function of the effective refractive index of the defect~\cite{Yeh_1977,joannopoulos2011photonic}.
It follows naturally that, for a properly designed structure terminated with an optically
anisotropic defect, the TE and TM dispersion relations can superimpose in the
$(k_{\parallel},\omega)$ space within the bandstructure forbidden region, where $k_{\parallel}$ is
the wavevector component parallel to the crystal surface. This eventually enables the coherent
excitation of both TE and TM modes in a wide spectral range, thus obtaining a chiral surface wave
upon the introduction of a $\pi/2$ phase shift between them.

A practical implementation of the proposed design is illustrated in Fig.~\ref{fig:1}. The
semi-infinite 1DPC consists of alternating  high ($n_{\mathrm{H}}$) and low ($n_{\mathrm{L}}$)
refractive index materials. For our structure we choose $\mathrm{Ta}_{2}\mathrm{O}_{5}$
($n_{\mathrm{H}}=2.06+0.001i$) and $\mathrm{SiO}_{2}$ ($n_{\mathrm{L}}=1.454+0.0001i$), which are a
standard choice for 1DPC fabrication, and allow for the generation of both TE and TM surface modes
~\cite{palik_1985,Kim_1996,Sinibaldi_2014}. Material dispersion is ignored for simplicity, yet it
can be straightforwardly included in the design process if needed. The 1DPC is then terminated with
an optically anisotropic defect characterized by the diagonal dielectric tensor:
\begin{equation}
  \bar{\bar{\varepsilon}}=
  \left(
   \begin{matrix}
    \varepsilon_{\mathrm{xx}} & 0 & 0 \\
    0 & \varepsilon_{\mathrm{xx}} & 0 \\
    0 & 0 & \varepsilon_{\mathrm{zz}}
   \end{matrix}
  \right).
\label{eq:tensor}
\end{equation}
In order to obtain the necessary uniaxial birefringence, we exploit form anisotropy and design the
defect as an additional $\mathrm{Ta}_{2}\mathrm{O}_{5}/\mathrm{SiO}_{2}$ multilayer with a period
$d_{\mathrm{def}}$ much smaller than the principal period $d_{\mathrm{1DPC}}$. The diagonal elements
of the tensor are then obtained by standard Maxwell homogenization and expressed
as~\cite{Andryieuski_2015}:
\begin{eqnarray}
&\varepsilon_{\mathrm{xx}}=f_{\mathrm{def}} \,\varepsilon_{\mathrm{H}} + (1-f_{\mathrm{def}}) \, \varepsilon_{\mathrm{L}},
\\
&\varepsilon_{\mathrm{zz}} = \left[ \frac{f_{\mathrm{def}}}{\varepsilon_{\mathrm{H}}} + \frac{(1-f_{\mathrm{def}})}{\varepsilon_{\mathrm{L}}}  \right]^{-1},
\label{eq:tensor_comp}
\end{eqnarray}
where $\varepsilon_{\mathrm{H,L}}=n_{\mathrm{H,L}}^{2}$, $f_{\mathrm{def}}$ is the filling factor of
the high refractive index material, and the upper surface of the defect is in contact with water.

\begin{figure}[t!]
\includegraphics{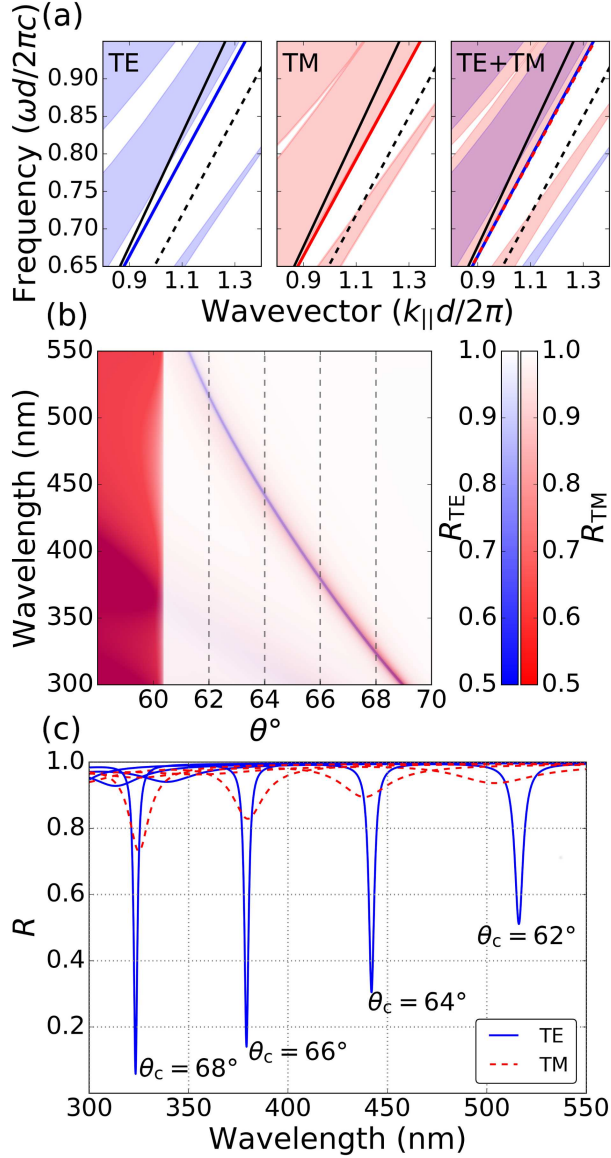}
\caption{\label{fig:2} (Color online) (a) 1DPC bandstructures for TE and TM illumination, and their
superposition. The black continuous and dashed lines represent the light line in water and in the
$n_{\mathrm{inc}}=1.53$ incident medium respectively. (b) Superimposed reflection maps for TE and TM
illuminations. The thin dashed lines indicate a spectral slice taken at constant incident angle
$\theta_{\mathrm{c}}$. (c) Spectral slices of the
reflectivity map for TE and TM illuminations at different incident angles $\theta_{\mathrm{c}}$.}
\end{figure}
The TE and TM bandstructure for the semi-infinite 1DPC, calculated with the MIT Photonic-Bands
package adopting the supercell method~\cite{Johnson2001_mpb}, are reported in Fig.~\ref{fig:2}(a).
The 1DPC supports both TE and TM surface modes, highlighted by the lines within the forbidden
regions of the band structure (white areas). The shaded regions correspond to allowed optical modes
extended inside the 1DPC. The same modes can either be extended or exponentially decaying in the
upper semi-space depending on their position above or below the light line $\omega=v k_{\parallel}$,
where $v=c/n_{\mathrm{water}}$ is the speed of light in water and $n_{\mathrm{water}}=1.33$. By
superimposing the two bandstructures, as displayed in the right panel of Fig.~\ref{fig:2}(a), the
optically anisotropic defect allows to obtain nicely overlapping TE and TM surface mode dispersion
relations in the $(k_{\parallel},\omega)$ space. This achievement offers the unique possibility to
launch surface modes with arbitrary polarization states in a wide spectral range, in direct contrast
to both standard surface plasmon polaritons and isolated Bloch surface waves.

We model the same phenomenon for an actual 1DPC with a finite number of periods, illuminated in
Kretschmann configuration for proper momentum matching. We illuminate the finite 1DPC, consisting of
2 crystal periods plus the anisotropic surface defect, from an incident medium with refractive index
$n_{\mathrm{inc}}=1.53$. The reflection maps for both TE and TM illuminations in
Fig.~\ref{fig:2}(b) display the surface modes as a narrow dark band (TE mode), superimposed to a lighter halo (TM mode), in agreement with the overlapping observed for the semi-infinite
1DPC, demonstrating how the adopted design approach is sound even for realistic finite size
structures. The total internal reflection onset is clearly visible around the $\theta\simeq60^{\circ}$ incident angle as a sharp variation in the reflected intensity.

The mode overlapping is also clearly illustrated in Fig.~\ref{fig:2}(c), which reports spectral
slices of the reflectivity maps computed at several coupling angles $\theta_{\mathrm{c}}$ (dashed
lines in Fig.~\ref{fig:2}(b)). The modes appear as well superimposed reflectivity dips at
coupling wavelengths $\lambda_{\mathrm{c}}$ ranging from 300~nm to 550~nm. In this framework,
the maximum optical chirality $C$ is obtained by illuminating the sample at
($\lambda_{\mathrm{c}}$,$\theta_{\mathrm{c}}$) with L or R elliptically polarized light. The
dephasing between the TE and TM incident plane wave components is thus defined as
$\phi_{\mathrm{c}}=\pi/2-\Delta\phi_{\mathrm{TE,TM}}$, where $\Delta\phi_{\mathrm{TE,TM}}$ is the
relative phase delay introduced by the 1DPC.
\begin{figure}[t!]
\includegraphics{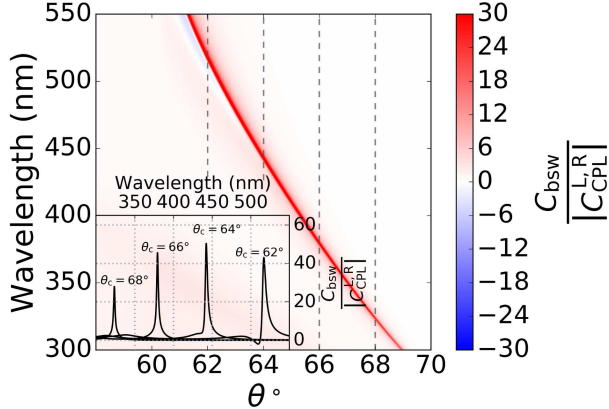}
\caption{\label{fig:3} (Color online) Optical chirality enhancement map calculated at the coupling
phase $\phi_{\mathrm{c}}$, 5~nm above the 1DPC surface. The thin dashed lines indicate a spectral
slice taken at constant incident angle $\theta_{\mathrm{c}}$. Inset: optical chirality enhancement
spectral slices taken at different incident angles $\theta_{\mathrm{c}}$.}
\end{figure}

Fig.~\ref{fig:3} reports the optical chirality map calculated 5~nm above the 1DPC surface, reminding
that along the $z$ axis we have an $e^{-k_{z}z}$ exponential decay with
$k_{z}=\sqrt{k_{\parallel}^{2}-k^{2}}$. We adopt a  unitary intensity L elliptical illumination
and normalize the results to the modulus of the optical chirality for a circularly polarized plane
wave $|C^{\mathrm{L,R}}_{\mathrm{CPL}}|$. The coherent superposition of the TE and TM surface modes
with the appropriate coupling phase $\phi_{\mathrm{c}}$ results in a sharp optical chirality
enhancement well above one order of magnitude in the whole analyzed spectral range. Superchirality
is only achieved if both TE and TM surface modes are simultaneously launched, while the mere
excitation of TE and TM evanescent waves is not sufficient to generate sizable superchirality.

The obtained superchiral field meets all the criteria for an ideal sensing platform. First, the
in-plane translational invariance of the system implies that the chirality enhancement is obtained
over arbitrarily large areas, where the only realistic constraints are imposed by limitations in the
illumination and fabrication processes. Second, the handedness of the optical chirality can be
readily switched by alternating between L and R incident elliptical polarization states.
Third, as illustrated in Fig.~\ref{fig:3}, the platform can operate in a wide spectral range in the
blue and near-UV end of the spectrum, opening up the possibility to perform surface enhanced CD
spectroscopy. Finally, it is worth stressing that the obtained surface wave is intrinsically chiral,
i.e. the optical chirality does not originate from the interference between the incident and local
fields, which is often the case for most plasmonic platforms~\cite{Nesterov_2016}.

To further investigate this design, we perform CD numerical simulations by adding a thin film of a
randomly oriented chiral analyte on top of the 1DPC. To this end we solve the multilayer transfer
matrix problem including the constitutive relations for the chiral medium~\cite{Jaggard_1992}:
\begin{eqnarray}
&\mathbf{D}=\varepsilon_{\mathrm{c}} \varepsilon_{0}\mathbf{E} + i\xi_{\mathrm{c}}\mathbf{B},
\\
&\mathbf{H}=\frac{\mathbf{B}}{\mu \mu{0}} + i\xi_{\mathrm{c}}\mathbf{E},
\label{eq:const_rel}
\end{eqnarray}
where $\varepsilon_{\mathrm{c}}$ and $\mu$ denote  the relative permittivity and permeability, and
$\xi_{\mathrm{c}}$ is the chiral optical admittance, i.e. the parameter that controls the chirality
strength~\cite{Jaggard_1992}. We work with a non-magnetic medium ($\mu=1$), and we define the optical parameters of the chiral medium as~\cite{Govorov_2012}:
\begin{eqnarray}
&\varepsilon_{\mathrm{c}}=\varepsilon_{\mathrm{c}0}-\gamma_{\mathrm{c}}
\left( \frac{1}{\hbar\omega-\hbar\omega_{0}+i\Gamma} -\frac{1}{\hbar\omega+\hbar\omega_{0}+i\Gamma}
\right),
\\
&\xi_{\mathrm{c}}=\beta_{\mathrm{c}}
\left( \frac{1}{\hbar\omega+\hbar\omega_{0}+i\Gamma} +\frac{1}{\hbar\omega-\hbar\omega_{0}+i\Gamma}
\right),
\label{eq:chi_mat}
\end{eqnarray}
where the coefficients $\gamma_{\mathrm{c}}$ and $\beta_{\mathrm{c}}$ determine the magnitude of the
absorptive and chiral properties. The energy of the molecular absorption resonance is
$\omega_{0}=2\pi c/ \lambda_{0}$, with $\lambda_{0}=380$~nm, while $\gamma_{\mathrm{c}}=2.5\cdot10^{-3}$eV
and $\beta_{\mathrm{c}}=2.5 \cdot 10^{-9} \mathrm{eV/\Omega}$ are chosen to match the optical
properties of Flavin mononucleotide (FMN)~\cite{Abdulrahman_2012}. Finally we set the damping and
background optical constants as $\Gamma=0.1$~eV and
$\varepsilon_{\mathrm{c}0}=1.77$~\cite{Nesterov_2016}.
\begin{figure}[t!]
\includegraphics{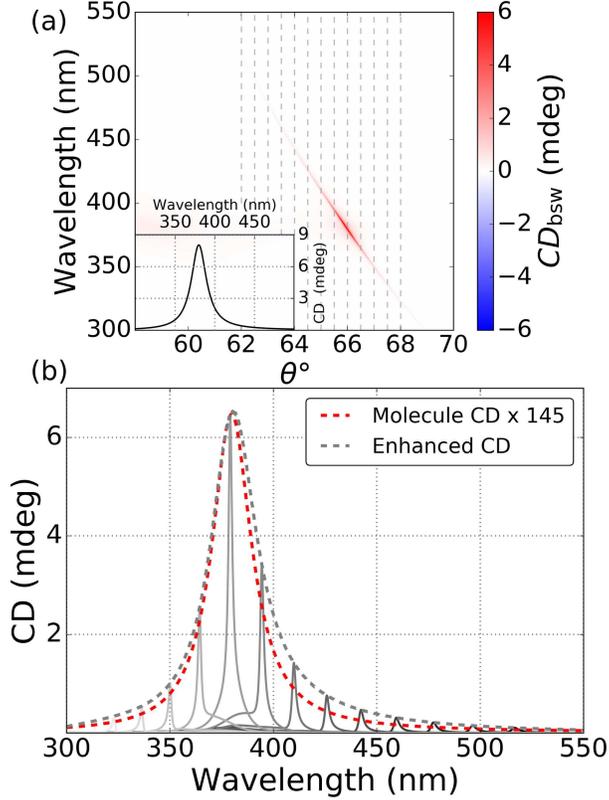}
\caption{\label{fig:4} (Color online) (a) Chirality map for a 100~ng/cm$^{2}$ FMN film deposited on
the 1DPC platform. The thin dashed lines indicate a spectral slice taken at constant incident angle
$\theta_{\mathrm{c}}$. Inset: CD spectrum for a  18~$\mu$g/cm$^{2}$ reference FMN film deposited on
a glass substrate. (b) Spectral slices taken at different incident angles $\theta_{\mathrm{c}}$
(gray continuous lines), envelope curve (gray dashed line) and CD spectrum for a  100~ng/cm$^{2}$
reference FMN film deposited on a glass substrate (red dashed line).}
\end{figure}

In the following we define the plane-wave reference CD signal as
$CD_{\mathrm{ref}}=A^{\mathrm{R}}_{\mathrm{ref}}-A^{\mathrm{L}}_{\mathrm{ref}}$, where
$A^{\mathrm{R,L}}_{\mathrm{ref}}=-\log(T^{\mathrm{R,L}})$ and the transmittance $T^{\mathrm{R,L}}$
is computed by normally illuminating a chiral thin film, deposited on a $n=1.45$ glass substrate,
with circularly polarized light of unitary intensity. On the other hand, we define the surface wave
CD signal as $CD_{\mathrm{CSW}}=A^{\mathrm{R}}_{\mathrm{CSW}}-A^{\mathrm{L}}_{\mathrm{CSW}}$. In
this case we use an equivalent absorbance defined as
$A^{\mathrm{R,L}}_{\mathrm{CSW}}=-\log(R^{\mathrm{R,L}})$, where the transmittance is substituted by
the reflectivity $R^{\mathrm{R,L}}$ in total internal reflection configuration, adopting L and R
incident elliptical polarization states. The CD signal enhancement is finally defined as
$CD_{\mathrm{CSW}}/CD_{\mathrm{ref}}$. The same quantities  can be used to define the reference and
surface wave dissymmetry factors as
$g_{\mathrm{ref}}=\frac{T^{\mathrm{R}}-T^{\mathrm{L}}}{T^{\mathrm{R}}+T^{\mathrm{L}}}$ and
$g_{\mathrm{CSW}}=\frac{R^{\mathrm{R}}-R^{\mathrm{L}}}{R^{\mathrm{R}}+R^{\mathrm{L}}}$,
respectively~\cite{Tang_2010}. Likewise the dissymmetry enhancement can be written as
$g_{\mathrm{CSW}}/g_{\mathrm{ref}}$. In the following we shall only refer to the CD enhancement
figure of merit, since it is possible to show that, for CD signal amplitudes that are typically very
small ($\ll 1$), $CD_{\mathrm{CSW}}/CD_{\mathrm{ref}} \simeq g_{\mathrm{CSW}}/g_{\mathrm{ref}}$.

Fig.~\ref{fig:4}(a) reports the CD signal map for a 5~nm chiral film, corresponding to a FMN surface
concentration of about 100~ng/cm$^{2}$~\cite{Abdulrahman_2012}. The map reveals a strong CD signal
perfectly aligned with the chiral surface mode. The maximum CD signal appears at the crossing
between the molecule electronic transition (inset of Fig.~\ref{fig:4}(a)) and the chiral modes,
reaching values up to $CD_{\mathrm{CSW}}\simeq6$~mdeg (millidegrees)\footnote{Historically, CD is
expressed as ellipticity in mdeg with $\Delta A_{\mathrm{mdeg}}= 32980 \cdot\Delta A$}, with
signal enhancements up to $CD_{\mathrm{CSW}}/CD_{\mathrm{ref}} \simeq 150$. In a traditional
transmittance setup the overall signal for the same molecular film would be well below the 0.1~mdeg
level, whereas 1~mdeg is considered the common limit of detection for standard CD
spectrometers~\cite{Nesterov_2016}. It follows that a very rough estimate of the limit of detection
for FMN films measured with the 1DPC returns a minimum measurable surface concentration of the order
of 15~ng/cm$^{2}$. Therefore, we emphasize that, even in the case of extremely low concentrations
($\simeq$15~ng/cm$^{2}$), we can directly detect the absorption CD signal without recurring to
fluorescence measurements, at variance with previous superchiral
approaches~\cite{Tang_2010,Tang_2011}.

It is finally of the utmost importance to verify that the wide spectral range of the chiral sensing
platforms allows for the recostruction of CD spectral lineshapes. This is simply accomplished by
performing several CD spectral slices at different coupling angles $\theta_{\mathrm{c}}$ (dashed
lines in Fig.~\ref{fig:4}(a)), and then calculating the corresponding envelope curve.
Fig.~\ref{fig:4}(b) demonstrates how the sensor provides CD signal enhancement above 2 orders of
magnitude, and how the envelope curve procedure allows to faithfully reconstruct the molecule CD
lineshape, thus laying the foundations for surface enhanced CD spectroscopy measurements. It is
nevertheless worth noting that this approach introduces a minor overestimation of the transition
linewidth, caused by a slight modulation in the sensor efficiency attributed to the molecule
resonant absorption.

In conclusion, we have presented the design of a 1DPC sensing platform capable of supporting chiral
surface waves generated by the coherent superposition of TE and TM surface modes. Chiral surface
waves provide homogeneous, superchiral and switchable local fields over arbitrarily large areas in a
wide spectral range. The present design provides optical chirality enhancements well above 1 order
of magnitude and CD signal enhancements of more than 2 orders of magnitude, allowing for the direct
measurement of absorption CD signals down to surface concentrations of the order of 15~ng/cm$^{2}$,
hence paving the way to fluorescence-free surface enhanced CD sensing and spectroscopy on extremely
low analyte amounts. The simple geometry makes the system an ideal candidate for automated design,
optimization procedures and straightforward fabrication with standard multilayer deposition
techniques. Most importantly, the adopted geometry is ideal for the integration with lab-on-chip
technologies and microfluidic networks on top of the surface of the device, allowing for the
targeting of molecules in solutions in well controlled environments, extremely small volumes and low
concentrations. Finally, the planar geometry and the employed materials are also ideal for surface
functionalization and for the analysis of iso-oriented chiral molecules, therefore allowing, as a
notable example, for the analysis of the spectral signatures from tertiary and quaternary structures
in proteins~\cite{Tullius_2015}.

\bibliography{bibliography.bib}

\end{document}